\documentclass[a4paper,11pt]{article}
\usepackage{jheppub}
\usepackage{array}
\usepackage{amsmath,amssymb,latexsym}
\usepackage{bm}
\usepackage[svgnames]{xcolor}

\usepackage{graphicx}
\usepackage{float}
\usepackage{subcaption}

\usepackage{mathtools}
\usepackage{physics}
\usepackage{braket}

\usepackage{natbib}
\usepackage{graphicx}

\usepackage{shuffle} 
\usepackage{slashed} 

\usepackage{multirow}

\usepackage[normalem]{ulem}
\newcommand{\stkout}[1]{\ifmmode\text{\sout{\ensuremath{#1}}}\else\sout{#1}\fi}

\allowdisplaybreaks

\usepackage{amsthm}

\begin{document}
	
	\title{NLO QCD sum rules analysis of $1^{-+}$ tetraquark states}
	
	\author{Wei-Yang Lai,}
	\author{Hong-Ying Jin}
	\affiliation{Zhejiang Institute of Modern Physics, School of Physics, Zhejiang University, Hangzhou 310027, China}
	\emailAdd{laiweiyang@zju.edu.cn}
	\emailAdd{jinhongying@zju.edu.cn}

	\abstract{We present a next-to-leading order QCD sum rule analysis of $J^{PC}=1^{-+}$ light tetraquark states. By investigating various compact tetraquark and molecular configurations, we determine the mass spectrum of these states. Our calculations exclude the possibility that $\pi_{1}(1400)$ is a tetraquark or hybrid-tetraquark mixture. This suggests that it may not exist, which is consistent with recent experimental results. In contrast, we obtained multiple $1^{-+}$ states around $2.0\,\text{GeV}$ that match well with $\pi_{1}(2015)$, which makes us confident that $\pi_{1}(2015)$ is a tetraquark candidate. As for $\pi_{1}(1600)$, our results indicate that the tetraquark currents tend to couple to heavier states, reducing the possibility of it being a tetraquark, while earlier studies suggested the opposite.
}
	

	\maketitle
	
	\newpage

	\section{Introduction}
	The study of non-$q\bar{q}$ mesons beyond the conventional quark model remains a central topic in hadron physics\cite{GELLMANN1964214,Albuquerque_2019}. QCD predicts the existence of mesons with various structures, such as hybrid states containing excited gluons\cite{li2025revisingmasslighthybrid,PhysRevD.107.034010}, glueballs composed purely of gluons\cite{adsfwefcv}, and multiquark states\cite{CPC:10.1088/1674-1137/ac2a1d,hv4x-dnmt,CPC:10.1088/1674-1137/43/3/034104,PhysRevD.91.054034,Dong2020,CPC:10.1088/1674-1137/ac0b3b}. Mesons with the exotic quantum numbers $J^{PC}=1^{-+}$ have attracted much attention and are typically regarded as candidates for hybrids, tetraquarks, or their mixtures. In particular, the ambiguity surrounding the isovector state $\pi_1(1400)$ has persisted for a long time. $\pi_1(1400)$ appeared in the $\pi^- p \to \pi^0 \eta n$ reaction\cite{ALDE1988397}, and it was later observed in the $\pi^- p \to \eta \pi^- p$ channel\cite{PhysRevLett.79.1630,PhysRevD.60.092001}. However, some recent evidence suggests that $\pi_{1}(1400)$ may be an artifact of the $\pi_{1}(1600)$ particle and might not exist\cite{Kopf2021}. In fact, a single pole is sufficient to fit the experimental data for both $\pi_{1}(1400)$ and $\pi_{1}(1600)$. In Ref.\cite{PhysRevD.71.011502}, $\pi_{1}(1400)$ was interpreted as a molecular tetraquark, while Ref.\cite{NARISON2009319} suggested that a hybrid-tetraquark mixture might explain $\pi_{1}(1400)$, a conclusion also reached in later work\cite{PhysRevD.105.054030}. Studies on hybrids indicate that the mass of a $1^{-+}$ hybrid state is higher than $1.7\,\text{GeV}$\cite{PhysRevD.52.5242,MEYER201521,Huang2015}, making it difficult to match the mass of $\pi_{1}(1400)$. Therefore, within the framework of QCD sum rules, the existence of $\pi_{1}(1400)$ depends on whether there exists a tetraquark state with a mass close to or less than $1.4\,\text{GeV}$.
	
	In Ref.\cite{PhysRevD.78.054017}, a series of compact tetraquark currents were calculated, and several isospin-1 $1^{-+}$ compact tetraquark states with masses around 1.6 GeV and 2.0 GeV were obtained. However, all the studies mentioned above on tetraquark states were only calculated to Leading Order (LO) without considering next-to-leading order (NLO) contributions, which can sometimes be significant\cite{li2025revisingmasslighthybrid}. In this work, we present the first NLO analysis of the $1^{-+}$ tetraquark states. Using the latest phenomenological parameters and a more precise running coupling, we reanalyze the possibility of the existence of $\pi_1(1400)$.

	\section{Tetraquark Currents and Renormalization}
	In past studies, some currents involving only the lighter $u$, $d$, and $s$ quarks and their antiquarks yielded lower ground-state masses in the spectral density when calculated at LO\cite{PhysRevD.78.054017, PhysRevD.71.011502}. Therefore, they are suitable subjects for this study, such that after adding NLO corrections, their masses might approach $1.4\,\text{GeV}$. For compact tetraquark currents, constructed by quark pairs $(qq)$ and antiquark pairs $(\bar{q}\bar{q})$, considering charge conjugation symmetry and flavor structure, we have 4 compact tetraquark currents $\eta_1 ^\mu \sim \eta_4^\mu$:
	\begin{equation}\label{2.3.1}
		\begin{split}
			\eta_1^\mu&=u_a^T C\gamma^\mu d_b (\bar{u}_aC\bar{d}^{\,T}_b + \bar{u}_bC\bar{d}^{\,T}_a) + u_a^T C d_b (\bar{u}_a\gamma^\mu C\bar{d}^{\,T}_b + \bar{u}_b\gamma^\mu C\bar{d}^{\,T}_a),\\
			\eta_2^\mu&=u_a^T C\sigma^{\mu\nu}\gamma^5 d_b (\bar{u}_a\gamma_\nu\gamma^5C\bar{d}^{\,T}_b + \bar{u}_b\gamma_\nu\gamma^5C\bar{d}^{\,T}_a) + u_a^T C \gamma_\nu\gamma^5d_b (\bar{u}_a\sigma^{\mu\nu}\gamma^5C\bar{d}^{\,T}_b + \bar{u}_b\sigma^{\mu\nu}\gamma^5 C\bar{d}^{\,T}_a),\\
			\eta_3^\mu&=u_a^T C\gamma^\mu d_b (\bar{u}_aC\bar{d}^{\,T}_b - \bar{u}_bC\bar{d}^{\,T}_a) + u_a^T C d_b (\bar{u}_a\gamma^\mu C\bar{d}^{\,T}_b - \bar{u}_b\gamma^\mu C\bar{d}^{\,T}_a),\\
			\eta_4^\mu&=u_a^T C\sigma^{\mu\nu}\gamma^5 d_b (\bar{u}_a\gamma_\nu\gamma^5C\bar{d}^{\,T}_b - \bar{u}_b\gamma_\nu\gamma^5C\bar{d}^{\,T}_a) + u_a^T C \gamma_\nu\gamma^5d_b (\bar{u}_a\sigma^{\mu\nu}\gamma^5C\bar{d}^{\,T}_b - \bar{u}_b\sigma^{\mu\nu}\gamma^5 C\bar{d}^{\,T}_a),\\
		\end{split}
	\end{equation}
	where $C$ is the charge conjugation operator, and $\sigma^{\mu\nu} = \frac{i}{2} [\gamma^{\mu}, 
	\gamma^{\nu}]$. If the $d$ quark is replaced by an $s$ quark, we obtain
	
	\begin{equation}
		\begin{split}
			\eta_5^\mu&=u_a^T C\gamma^\mu s_b (\bar{u}_aC\bar{s}^T_b + \bar{u}_bC\bar{s}^T_a) + u_a^T C s_b (\bar{u}_a\gamma^\mu C\bar{s}^T_b + \bar{u}_b\gamma^\mu C\bar{s}^T_a),\\
			\eta_6^\mu&=u_a^T C\sigma^{\mu\nu}\gamma^5 s_b (\bar{u}_a\gamma_\nu\gamma^5C\bar{s}^T_b + \bar{u}_b\gamma_\nu\gamma^5C\bar{s}^T_a) + u_a^T C \gamma_\nu\gamma^5s_b (\bar{u}_a\sigma^{\mu\nu}\gamma^5C\bar{s}^T_b + \bar{u}_b\sigma^{\mu\nu}\gamma^5 C\bar{s}^T_a),\\
			\eta_7^\mu&=u_a^T C\gamma^\mu s_b (\bar{u}_aC\bar{s}^T_b - \bar{u}_bC\bar{s}^T_a) + u_a^T C s_b (\bar{u}_a\gamma^\mu C\bar{s}^T_b - \bar{u}_b\gamma^\mu C\bar{s}^T_a),\\
			\eta_8^\mu&=u_a^T C\sigma^{\mu\nu}\gamma^5 s_b (\bar{u}_a\gamma_\nu\gamma^5C\bar{s}^T_b - \bar{u}_b\gamma_\nu\gamma^5C\bar{s}^T_a) + u_a^T C \gamma_\nu\gamma^5s_b (\bar{u}_a\sigma^{\mu\nu}\gamma^5C\bar{s}^T_b - \bar{u}_b\sigma^{\mu\nu}\gamma^5 C\bar{s}^T_a).
		\end{split}
		\label{tetra_us}	
	\end{equation}
	Furthermore, for molecular tetraquark states, we also consider the following two currents\cite{PhysRevD.71.011502}
	\begin{align}
		J_1^\mu&=\frac{1}{2}\left(\bar{u}\gamma^5 u-\bar{d}\gamma^5d\right)\left(\bar{u}\gamma^5\gamma^\mu  u+\bar{d}\gamma^5\gamma^\mu d\right),\\
		J_2^{\mu\nu}&=\epsilon^{\mu\nu\rho\sigma}\left(\bar{u}\gamma^5\gamma_\rho d \,\bar{d}\gamma_\sigma u - \bar{d}\gamma^5\gamma_\rho u \,\bar{u}\gamma_\sigma d\right),
	\end{align}
	where $\epsilon^{\mu\nu\rho\sigma}$ is the totally antisymmetric tensor, with the convention $\epsilon^{0123}=+1$.

	When expanding the perturbative part of the correlation functions for these tetraquark currents to NLO, we need to calculate the Feynman diagrams as shown in Fig.\ref{ope_d00}.
	Since tetraquark currents are composite operators, the calculation results will contain non-local divergence terms such as
	\begin{equation}
		\log(-q^2/\mu^2)/\varepsilon.
	\end{equation}
	These non-local divergences cannot be completely removed by conventional Lagrangian renormalization methods; the operator currents themselves must be renormalized. For operator currents $\eta_1^\mu$ and $\eta_3^\mu$, using the tetraquark renormalization method in Ref.\cite{li2025qcdsumruleanalysis}, with the Feynman gauge, we can obtain
	\begin{align}
		\left(u_a^T C d_b \,\bar{u}_a\gamma^\mu C\bar{d}^{\,T}_b\right)_r=&\left(Z_2^{-2}-\frac{2C_A^2+1 }{32\pi^2\varepsilon C_A}g^2\right)u_a^T C d_b\,\bar{u}_a\gamma^\mu C\bar{d}^{\,T}_b  +  \frac{3}{32\pi^2\varepsilon}g^2u_a^T C d_b\,\bar{u}_b\gamma^\mu C\bar{d}^{\,T}_a\notag\\
		&-\frac{C_A i g^2}{32\pi^2\varepsilon}u_a^T C\sigma^{\mu\nu}\gamma^5 d_b \,\bar{u}_a\gamma_\nu\gamma^5C\bar{d}^{\,T}_b  +  \frac{i g^2}{32\pi^2\varepsilon}u_a^T C\sigma^{\mu\nu}\gamma^5 d_b \,\bar{u}_b\gamma_\nu\gamma^5C\bar{d}^{\,T}_a
		,\label{ren_tetra_1_a}\\
		\left(u_a^T C d_b \,\bar{u}_b\gamma^\mu C\bar{d}^{\,T}_a\right)_r=&\left(Z_2^{-2}-\frac{2C_A^2+1 }{32\pi^2\varepsilon C_A}g^2\right)u_a^T C d_b\,\bar{u}_b\gamma^\mu C\bar{d}^{\,T}_a  +  \frac{3}{32\pi^2\varepsilon}g^2u_a^T C d_b\,\bar{u}_a\gamma^\mu C\bar{d}^{\,T}_b\notag\\
		&+\frac{C_A i g^2}{32\pi^2\varepsilon}u_a^T C\sigma^{\mu\nu}\gamma^5 d_b \,\bar{u}_b\gamma_\nu\gamma^5C\bar{d}^{\,T}_a  -  \frac{i g^2}{32\pi^2\varepsilon}u_a^T C\sigma^{\mu\nu}\gamma^5 d_b \,\bar{u}_a\gamma_\nu\gamma^5C\bar{d}^{\,T}_b\notag\\
		&+\frac{1}{48\pi^2\varepsilon}\left(\bar{u}D^\alpha G_{\alpha\beta} \gamma^\beta\gamma^\mu\gamma^5 u - \bar{d}D^\alpha G_{\alpha\beta} \gamma^\beta\gamma^\mu\gamma^5 d\right)
		,\label{ren_tetra_1_b}
	\end{align}
	In the above equations,
	\begin{equation}
		Z_2=1-\frac{g^2 C_F}{16\pi^2\epsilon},
	\end{equation}
	where $C_F=4/3$ is the quadratic Casimir operator of the fundamental representation of the $SU(3)$ group, and $C_A=3$ is the quadratic Casimir operator of the adjoint representation of the $SU(3)$ group; these parameters are also known as the quark and gluon color factors, respectively. For convenience, we suppress the generator matrix $T$ of $SU(3)$ and the coupling constant $g$, i.e.,
	\begin{equation}
		D^\alpha G_{\alpha\beta} \equiv g D^\alpha G^n_{\alpha\beta} T^n.
	\end{equation}
	It can be seen that for the two different color index summation methods, $u_a^T C d_b \,\bar{u}_a\gamma^\mu C\bar{d}^{\,T}_b$ and $u_a^T C d_b \,\bar{u}_b\gamma^\mu C\bar{d}^{\,T}_a$, the renormalization results differ by a hybrid-like current $D^\alpha G_{\alpha\beta}$, which only appears in operator currents where the summation is over different quark color indices. By replacing the $d$ quark in Eqs.(\ref{ren_tetra_1_a}) and (\ref{ren_tetra_1_b}) with an $s$ quark, we obtain the operator renormalization results for tetraquark currents $\eta_5^\mu$ and $\eta_7^\mu$. As for the molecular tetraquark currents, they can be obtained through linear combinations of the following renormalized operator currents:

	\begin{align}
		\left(\bar{u}\gamma^5 u\,\bar{d}\gamma^5\gamma^\mu d\right)_r=&\left(Z_2^{-2}+\frac{5C_F}{16\pi^2\varepsilon}g^2\right)\bar{u}\gamma^5 u\,\bar{d}\gamma^5\gamma^\mu d + \frac{ig^2}{8\pi^2\varepsilon}\bar{u}T^n\sigma^{\mu\nu}u\,\bar{d}\gamma_\nu T^n d,\\
		\left(\bar{u}\gamma^5 u\,\bar{u}\gamma^5\gamma^\mu u\right)_r=&\left(Z_2^{-2}+\frac{5C_F}{16\pi^2\varepsilon}g^2\right)\bar{u}\gamma^5 u\,\bar{u}\gamma^5\gamma^\mu u + \frac{ig^2}{8\pi^2\varepsilon}\bar{u}T^n\sigma^{\mu\nu}u\,\bar{u}\gamma_\nu T^n u\notag\\
		&+\frac{i}{24\pi^2\varepsilon}\bar{u}D^\alpha G_{\alpha\beta}\sigma^{\beta\mu}u.
	\end{align}
	The renormalization of other operator currents is placed in Appendix \ref{apx2}.

	\section{QCD Sum Rules}
	
	\begin{figure}[h!]
		\centering
		\includegraphics[width=0.89\textwidth]{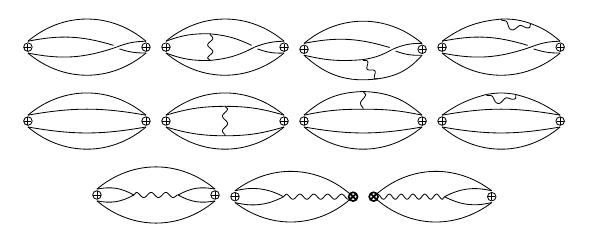}
		\caption{Feynman diagrams for the perturbative term of tetraquark currents. Permutation diagrams are omitted. The bold cross vertex represents the hybrid-like counterterm. For compact tetraquark currents, the first four diagrams are unnecessary.\label{ope_d00}}
	\end{figure}

	QCD sum rules\cite{2f559d0e1b0142098757851dd8a536ba,cohen1995qcd,NARISON202394} determine the corresponding hadronic spectral structure through the poles of operator current correlation functions in the complex plane. To calculate the correlation function, we first perform an Operator Product Expansion (OPE):
	\begin{equation}
		\Pi(q)=i\int d^4x e^{iqx}\langle 0|T\{J^\mu(x) J^\nu(0)\}|0\rangle = C_0(q) + C_4(q)\langle O^4\rangle + C_6(q)\langle O^6\rangle +\cdots,
	\end{equation}
	where $C_i(q)$ are Wilson coefficients, and $\langle O^i\rangle$ denote vacuum condensates(i.e., the vacuum expectation values of various local operators). The first Wilson coefficient $C_0(q)$ corresponds to the perturbative diagrams in Fig.\ref{ope_d00}, which accounts for the high-energy contribution. Subsequent coefficients like $C_4(q)$ and $C_6(q)$ can be calculated by cutting certain propagators in the perturbative diagrams or introducing zero-momentum gluons from the vacuum; they represent the contribution from the low-energy region. For this part of the contribution, it is suitable to calculate in coordinate space. We can use the techniques in Refs.\cite{PhysRevD.78.054017,Aliev_2024,SONG2019160,Di_2019} to handle these non-perturbative contributions. The quark propagator including vacuum condensates is
	\begin{equation}
		\begin{split}
			i S^{ab} &\equiv  \langle 0 |
			T [ q^a(x) \bar{q}^b(0) ] | 0 \rangle\\ 
			&= { i \delta^{ab} \over 2 \pi^2 x^4 } \slashed{x} - {ig \over
				32\pi^2}  G^n_{\mu\nu} T^{n\,ab}{1 \over x^2}
			(\sigma^{\mu\nu}.\slashed{x} + \slashed{x}.\sigma^{\mu\nu}) - { \delta^{ab}
				\over 12 } \langle \bar q q \rangle \\ 
			&\quad - { \delta^{ab} x^2 \over
				192 } \langle g \bar q \sigma G q \rangle - { m_q \delta^{ab}
				\over 4 \pi^2 x^2 } + { i \delta^{ab} m_q \langle \bar q q \rangle
				\over 48 }  \slashed{x} + { i \delta^{ab} m_q^2 \over 8 \pi^2 x^2 }
			\slashed{x}
		\end{split}
	\end{equation}
	
	The Lorentz structure of operator currents indicates that a general vector current can couple to both vector particles and scalar particles. Therefore, we need to isolate the part corresponding to the $1^{-+}$ vector particle in the correlation function. In momentum space, the correlation function of the vector current can be decomposed into
	\begin{equation}
		i\int d^4x e^{iqx}\langle 0|T\{J^\mu(x) J^\nu(0)\}|0\rangle = \left(\frac{q^\mu q^\nu}{q^2}-g^{\mu\nu}\right)\Pi^V(q) + g^{\mu\nu}\Pi^S(q),
	\end{equation}
	The term $\left(\frac{q^\mu q^\nu}{q^2}-g^{\mu\nu}\right)$ on the right side of the equation is the transverse polarization factor, and $g^{\mu\nu}$ is the longitudinal polarization factor. The vector particle couples to the vector current $J^\mu$ through the matrix element
	\begin{equation}
		\langle 0|J^\mu|V(q)\rangle=\epsilon^\mu f(q^2).
	\end{equation}
	Meanwhile, the polarization vector $\epsilon^\mu$ satisfies the completeness relation
	\begin{equation}\label{3.5}
		\sum_{\lambda=-1}^{1}\epsilon_{\lambda}^{\mu}\,\epsilon_{\lambda}^{\nu}=\frac{q^\mu q^\nu}{q^2}-g^{\mu\nu}.
	\end{equation}
	Therefore, the spectral density of the vector particle is related to the transverse polarization part of the correlation function. For vector currents $\eta_1^\mu \sim \eta_8^\mu$ and $J_1^\mu$, only the $\Pi^V(q)$ part is considered. For tensor currents, they can couple to scalar particles, vector particles, and tensor particles. Similar to the analysis above, considering that $J_2^{\mu\nu}$ is antisymmetric with respect to Lorentz indices, we decompose the tensor current operator correlation function in momentum space as
	\begin{equation}
		\begin{split}
			i\int d^4x e^{iqx}\langle 0|J_2^{\mu\nu}(x)\, J_2^{\dagger\,\alpha\beta}|0\rangle=&(\eta^{\mu\alpha}\eta^{\nu\beta}-\eta^{\nu\alpha}\eta^{\mu\beta})\Pi_{J_2}^{\prime V}(q^2)\\
			&+(\eta^{\mu\alpha}q^\nu q^\beta - \eta^{\nu\alpha}q^\mu q^\beta - \eta^{\mu\beta}q^\nu q^\alpha + \eta^{\nu\beta}q^\mu q^\alpha)\frac{1}{q^2}\Pi_{J_2}^V(q^2),
		\end{split}
	\end{equation}
	where $\eta^{\mu\nu}\equiv\frac{q^\mu q^\nu}{q^2}-g^{\mu\nu}$. The coupling rule for vector particles and tensor currents is
	\begin{equation}
		\langle 0|J_2^{\mu\nu}|V(q)\rangle=(q^\mu\epsilon^\nu - q^\nu\epsilon^\mu) f(q^2),
	\end{equation}
	Combined with Eq.(\ref{3.5}), it can be seen that the spectral density of the vector particle corresponds to the $\Pi_{J_2}^V(q^2)$ part.
	
	We use techniques such as dimensional regularization and the $\overline{\mathrm{MS}}$ subtraction scheme, adopting the BMHV scheme for $\gamma^5$, to obtain the OPE results of the correlation function. For example, the correlation function $\Pi^V$ of $\eta_1^\mu$ is
	
	\begin{equation}
		\begin{split}
			\Pi^V_{\eta_1}=&q^8\left[ \left(-\frac{79 g_s^2}{2654208 \pi ^8}-\frac{1}{18432 \pi ^6}\right) \log \left(-\frac{q^2}{\mu ^2}\right)-\frac{5 g_s^2 }{1327104 \pi ^8}\log ^2\left(-\frac{q^2}{\mu ^2}\right)\right] \\ 
			&+q^4\frac{ g_s^2  }{18432 \pi ^6}\log \left(-\frac{q^2}{\mu ^2}\right) \langle  GG \rangle+-q^2\frac{ 1}{18 \pi ^2}\log \left(-\frac{q^2}{\mu ^2}\right) \left\langle  \bar{q}q \right\rangle ^2\\ 
			&+\frac{1}{12 \pi ^2}\log \left(-\frac{q^2}{\mu ^2}\right) \left\langle  \bar{q}q \right\rangle  \left\langle  \bar{q}Gq \right\rangle +\frac{1}{q^2}\left(\frac{5 g_s^2 }{864 \pi ^2}\langle  GG \rangle  \left\langle  \bar{q}q \right\rangle ^2-\frac{1}{48 \pi ^2}\left\langle  \bar{q}Gq \right\rangle ^2\right)\\ 
			&+\frac{1}{q^4}\left(-\frac{g_s^2}{576 \pi ^2} \langle  
			GG \rangle  \left\langle  \bar{q}q \right\rangle  \left\langle  \bar{q}Gq \right\rangle -\frac{32}{81} g_s^2 \left\langle  \bar{q}q \right\rangle ^4\right).
			\label{c_eta_1}
		\end{split}
	\end{equation}	
	The remaining correlation functions are given in Appendix \ref{apx1}.
	
	Using the dispersion relation and adopting the ``$\delta$ + continuum'' assumption, we have
	\begin{equation}\label{3.9}
		\frac{1}{\pi}\text{Im}\Pi^V(q) = f^2\delta(s-m^2)+\theta(s-s_0)\rho_c(s),
	\end{equation}
	where $m$ is the mass of the lowest resonance state, $f$ is the coupling strength, $s_0$ is the continuum threshold, and $\rho_c(s)$ is the spectral density of the continuum states. Applying the Borel transform to both sides of the above equation yields the $n$-th moment $\mathcal{M}^n(\tau,s_0)$:
	\begin{equation}
		\mathcal{M}^n(\tau,s_0)\equiv\frac{1}{\pi}\int_0^{s_0}ds\ s^n e^{-\tau s}\, \text{Im}\Pi^V(s)=f^2 m^{2n}e^{-\tau m^2}.
		\label{moment}
	\end{equation}
	The mass $m$ of the lowest resonance state in the spectral density can be obtained from the ratio of the moment:
	\begin{equation}
		\mathcal{R}^n(\tau,s_0)\equiv\frac{\mathcal{M}^{n+1}(\tau,s_0)}{\mathcal{M}^n(\tau,s_0)}=m^2,
		\label{ratio1}
	\end{equation}
	By taking the ratio of moments, the coupling constant $f$ cancels out.

	\section{Numerical Analysis}
	In the numerical analysis, we adopt the following quark masses and vacuum condensate parametersat the scale $\mu = 2\,\text{GeV}$\cite{pdg,NARISON2015189}:
	\begin{align*}
		m_s &= 93.5\pm 0.8\text{MeV},\\
		\frac{g^2}{4\pi}\langle GG\rangle & = 0.07\pm0.02\,\text{GeV}^4,\\
		\langle \bar{q}q\rangle &= -(0.276)^3\,\text{GeV}^3,\\
		\langle\bar{s}s\rangle &= 0.74\langle \bar{q}q\rangle,\\ 
		\langle\bar{q}Gq\rangle\!
		&= M_0^2\langle \bar{q}q\rangle,\\
		\langle\bar{s}Gs\rangle &= M_0^2\langle \bar{s}s\rangle,\\
		M_0^2&=0.8\pm0.2\,\text{GeV}^2,
		\label{qcd_parameters}
	\end{align*}
	where $q=u \text{ or } d$, $\langle GG\rangle = \langle G^n_{\mu\nu}G^{n\mu\nu}\rangle$, $\langle \bar{q}Gq\rangle=\langle\bar{q}gT^nG^n_{\mu\nu}\sigma^{\mu\nu}q\rangle$, and the masses of light quarks $u$ and $d$ are neglected. The running coupling constant $\alpha_s$ is taken in the one-loop approximation:
	\begin{equation}
		\alpha_s(\mu^2)=\frac{\alpha_s(m^2_\tau)}{1+\frac{\beta_0}{4\pi}\alpha_s(m^2_\tau)\log\big(\frac{\mu^2}{m^2_\tau}\big)},
	\end{equation}
	where $m_\tau=1776.93\pm0.09\text{MeV}$, $\alpha_s(m^2_\tau)=0.314\pm0.014$, with $n_f=3$ and $\beta_0=9$. Renormalization-group improvement is achieved by setting $\mu^2=1/\tau$.

\begin{figure}[h!]
		\centering
		\includegraphics[width=8.4cm]{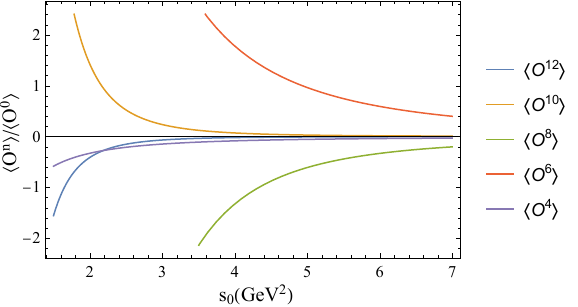}
		\caption{Ratio of condensate terms $\langle O^n\rangle$ to the perturbative term $\langle O^0\rangle$ for $\eta_1^{\mu}$, at $\tau=0.3~\mathrm{GeV}^{-2}$.}
		\label{con_per}
	\end{figure}

	 We calculated the relative contributions of the perturbative term and condensate terms to the moment $\mathcal{M}^0(\tau,s_0)$ of the current $\eta_1^{\mu}$, as shown in Fig.\ref{con_per}. It can be seen that the OPE convergence is well, so the calculation results are reliable.

\begin{figure}[h!]
		\centering
		\includegraphics[width=5.4cm]{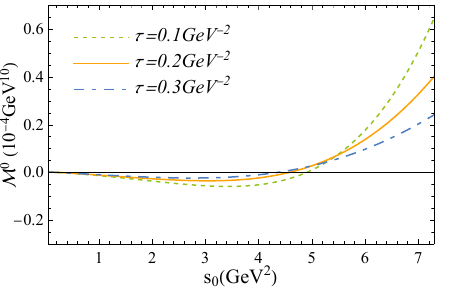}
		\hspace{1cm}
		\includegraphics[width=5.4cm]{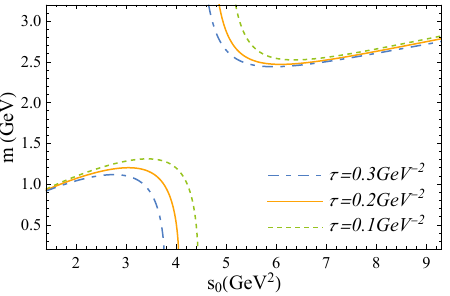}
		\caption{NLO results for the moment $\mathcal{M}^0$ and mass $m$ versus the $s_0$ for the current $J_2^{\mu\nu}$}
		\label{mol_2_s0}
	\end{figure}

	Using the ratio of moments in Eq.~(\ref{ratio1}), we extract the mass of the lowest resonance by $m=\sqrt{\mathcal{R}^n}(\tau)$, which depends on the Borel parameter $\tau$ and the continuum threshold $s_0$.
	Since $\tau$ is an auxiliary parameter, the extracted mass should exhibit a
	stable plateau with respect to $\tau$. We therefore determine the mass from the
	extremum (or the flattest region) of the $m(\tau)$ curve and fix the values of $s_0$ and $\tau$. It is worth noting that the moment $\mathcal{M}^0$ corresponds to the spectral density and thus is always positive. If a negative value appears, it indicates that the QCD sum rule analysis breaks, and the corresponding mass is non-physical. In Fig.\ref{mol_2_s0}, we list the $\mathcal{M}^0 \sim s_0$ and $m \sim s_0$ curves for the tensor current $J_2^{\mu\nu}$. In the left figure, the region $s_0\lesssim 4.5\,\text{GeV}^2$ is non-physical, so the corresponding mass $\lesssim 1.5\,\text{GeV}^2$ in the right figure is not credible; the true mass should be $\gtrsim2.3	\,\text{GeV}^2$. In the subsequent analysis, we discuss within the physical interval where $\mathcal{M}^0$ is positive.
	
\begin{figure}[h!]
		\centering
		\includegraphics[width=5.4cm]{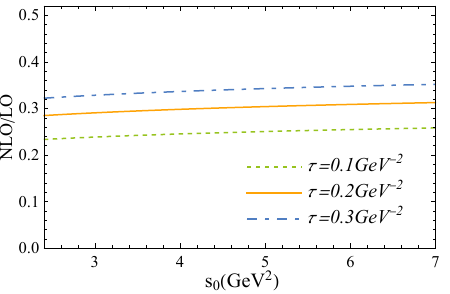}
		\hspace{1cm}
		\includegraphics[width=5.4cm]{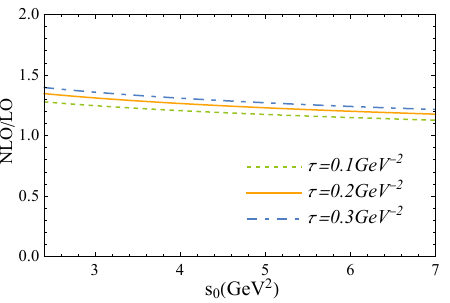}
		\caption{Ratio of NLO contribution to LO contribution in moment $\mathcal{M}^0$ for the currents $J_2^{\mu\nu}$(left) and $\eta_2^\mu$(right).}
		\label{nlo_lo_ratio}
	\end{figure}

\begin{figure}[h!]
		\centering
		\includegraphics[width=5.4cm]{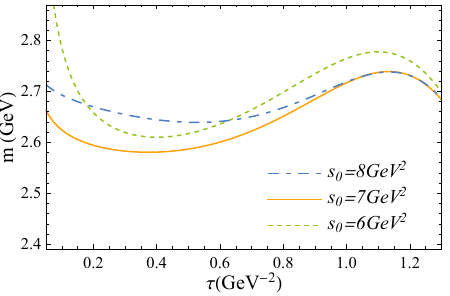}
		\hspace{1cm}
		\includegraphics[width=5.4cm]{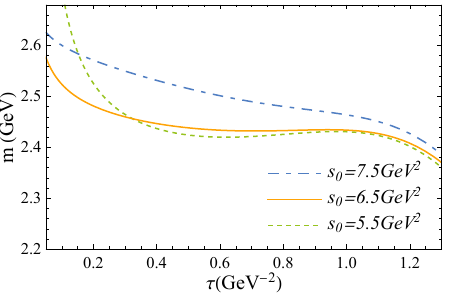}
		\caption{Mass predictions for the current $J_2^{\mu\nu}$ at LO(left) and NLO(right) levels.}
		\label{mol_2_tau}
	\end{figure}

	Fig.\ref{nlo_lo_ratio} shows that, for $J^{\mu\nu}_2$, the NLO perturbative correction is roughly $30\%$ of the LO contribution and thus cannot be neglected.For $\eta^\mu_2$, the NLO perturbative correction is even larger than the LO term, reaching about $130\%$ of the LO contribution. Furthermore, in Fig.\ref{mol_2_tau}, the left side shows the estimated mass of the operator current $J^{\mu\nu}_2$ when expanded to LO, and the right side shows the estimated mass after adding the NLO perturbation correction. Including the NLO correction significantly affects the mass determination. In particular, the mass curve is lowered by roughly $0.2\,\text{GeV}$ and becomes flatter, which leads to improved stability.


	
	In Appendix \ref{massfig}, we present the curves of the resonance mass $m$ as a function of the Borel parameter $\tau$ for all currents, with $s_0$ fixed at various values. Based on the stability criterion, we select three $s_0$ values that yield the flattest curves in each figure. The corresponding optimal mass estimates and $s_0$ values are listed in Table \ref{masstable}. The masses for the compact tetraquark currents $\eta_1^\mu \sim \eta_4^\mu$ (with $u$
	and $d$ quarks) are $2.0\,\text{GeV}$, $1.9\,\text{GeV}$, $1.7\,\text{GeV}$, and $2.0\,\text{GeV}$. For
	$\eta_5^\mu\sim\eta_8^\mu$ (with $u$ and $s$ quarks), the corresponding masses
	are $2.4\,\text{GeV}$, $2.0\,\text{GeV}$, $2.2\,\text{GeV}$, and $2.4\,\text{GeV}$. For the molecular currents, we
	obtain $1.8\,\text{GeV}$ for $J_1^\mu$ and $2.45\,\text{GeV}$ for
	$J_2^{\mu\nu}$. In comparison with previous LO QCD sum rule studies, our mass predictions show shifts of $0.1 \sim 0.3\,\text{GeV}$. These corrections are primarily attributed to the NLO contributions, as well as the updated QCD parameters, the use of a more precise running coupling, and renormalization group improvements.

	\begin{table}[h!]
		\centering
		\renewcommand{\arraystretch}{1.5}
		\begin{tabular*}{\textwidth}{@{\extracolsep{\fill}}lcccccccccc}
			\hline\hline
			& $\eta_1$ & $\eta_2$ & $\eta_3$ & $\eta_4$ & $\eta_5$ & $\eta_6$ & $\eta_7$ & $\eta_8$ & $J_1$ & $J_2$ \\ 
			\hline
			
			$m$   & 2.0 & 1.9 & 1.7 & 2.0 & 2.4 & 2.0 & 2.2 & 2.4 & 1.8 & 2.45\\ 
			$s_0$ & 5.0 & 4.5 & 3.5 & 5.0 & 6.5 & 5.0 & 6.0 & 6.5 & 4.0 & 6.5\\ 
			\hline\hline
		\end{tabular*}
		\caption{Resonance masses corresponding to operator currents}
		\label{masstable}
	\end{table}

	Within the framework of QCD sum rules, the $\pi_1(1600)$ and $\pi_1(2015)$ states remain compatible with the tetraquark interpretation. However, it is worth noting that our corrected results yield no $1^{-+}$ resonance mass corresponding to tetraquark currents arround or below $1.4\,\text{GeV}$. Consequently, the interpretation of the $\pi_1(1400)$ as a tetraquark or a hybrid-tetraquark mixture candidate is effectively excluded. Recent studies suggest that previous analyses of $\pi_1(1400)$ experimental data may be flawed. Upon reanalysis, a single resonance peak centered at $1.6\,\text{GeV}$ is sufficient to fit all experimental data. Therefore, it is concluded that the $\pi_1(1400)$ may not exist and its signal is merely an artifact of the $\pi_1(1600)$. Notably, the Particle Data Group (PDG) has included the $\pi_1(1400)$ data into the $\pi_1(1600)$ entry. Our calculation results provide further theoretical support for this perspective within the QCD sum rule framework.


	\newpage
	
	\appendix
	
	\section{Appendix}
	
	\subsection{$m(\tau)$ curves of resonance states}\label{massfig}
	
	\begin{figure}[h!]
		\centering
		\includegraphics[width=5.4cm]{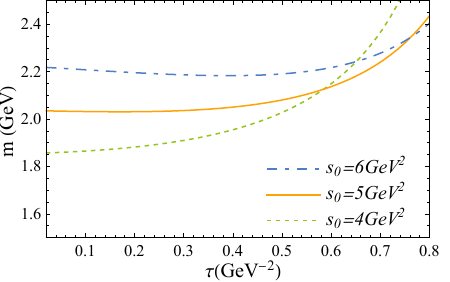}
		\hspace{1cm}
		\includegraphics[width=5.4cm]{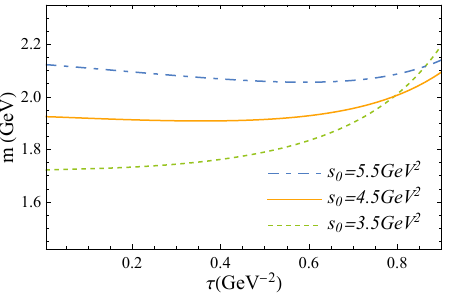} 
		\caption{Mass predictions for the current $\eta_1^\mu$(left) and $\eta_2^\mu$(right).}\label{eta_12}
	\end{figure}
	\begin{figure}[h!]
		\centering
		\includegraphics[width=5.4cm]{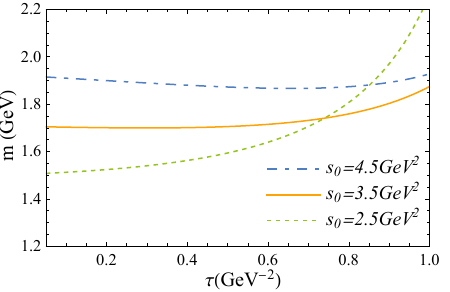}
		\hspace{1cm}
		\includegraphics[width=5.4cm]{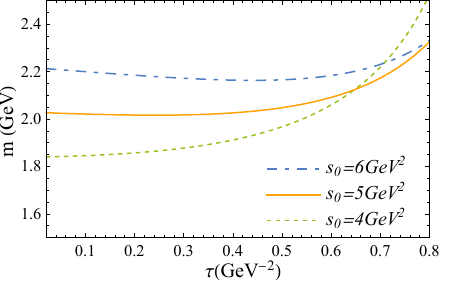} 
		\caption{Mass predictions for the current $\eta_3^\mu$(left) and $\eta_4^\mu$(right).}\label{eta_34}
	\end{figure}
	\begin{figure}[h!]
		\centering
		\includegraphics[width=5.4cm]{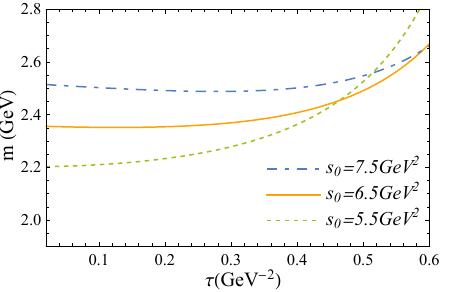}
		\hspace{1cm}
		\includegraphics[width=5.4cm]{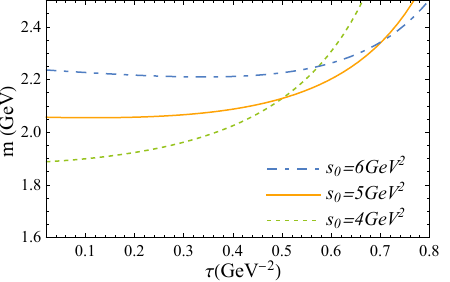} 
		\caption{Mass predictions for the current $\eta_5^\mu$(left) and $\eta_6^\mu$(right).}\label{eta_56}
	\end{figure}
	\begin{figure}[h!]
		\centering
		\includegraphics[width=5.4cm]{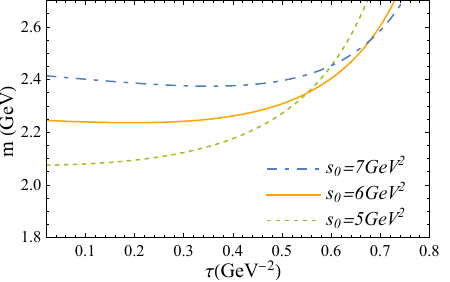}
		\hspace{1cm}
		\includegraphics[width=5.4cm]{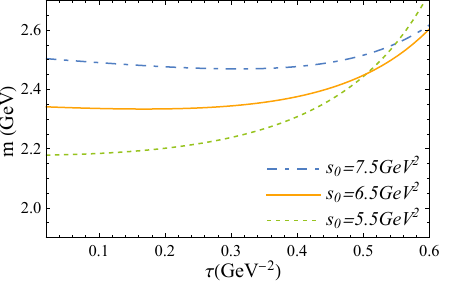} 
		\caption{Mass predictions for the current $\eta_7^\mu$(left) and $\eta_8^\mu$(right).}\label{eta_78}
	\end{figure}
	\begin{figure}[h!]
		\centering
		\includegraphics[width=5.4cm]{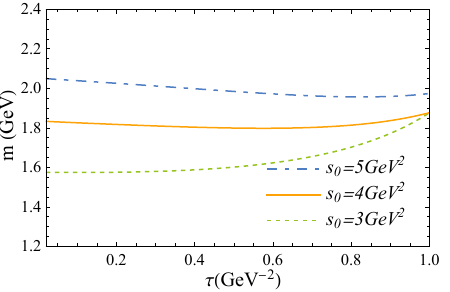}
		\hspace{1cm}
		\includegraphics[width=5.4cm]{./figs//mol_2_m_tau.pdf} 
		\caption{Mass predictions for the current $J_1^{\mu}$(left) and $J_2^{\mu\nu}$(right).}\label{mol_12}
	\end{figure}

	\subsection{Renormalized Operator Currents}\label{apx2}
	Renormalized operator current $J_2^{\mu\nu}$:
	\begin{equation}
		\begin{split}
			\left(\epsilon^{\mu\nu\rho\sigma}\,\bar{u}\gamma^5\gamma_\rho d\, \bar{d}\gamma_\sigma u\right)_r=\epsilon^{\mu\nu\rho\sigma}\,\bigg[&\left(Z_2^{-2}+\frac{5C_F}{8\pi^2\varepsilon}g^2\right)\bar{u}\gamma_\rho\gamma^5 d\, \bar{d}\gamma_\sigma u-\frac{g^2}{8\pi^2\varepsilon}\bar{u}\gamma^5\gamma_\rho 
			T^n d\, \bar{d}\gamma_\sigma\gamma^5  T^n u\\
			&-\frac{i}{48\pi^2\varepsilon}\epsilon_{\rho\sigma\beta\eta}\left(\bar{u}D_\alpha G^{\alpha\beta}\gamma^\eta u - \bar{d}D_\alpha G^{\alpha\beta}\gamma^\eta d\right)\bigg]
		\end{split}
	\end{equation}

	For $\eta_2^\mu$ and $\eta_4^\mu$, their renormalization can be obtained through linear combinations of the following renormalization currents:
	\begin{equation}
		\begin{split}
			\left(u_a^T C\sigma^{\mu\nu}\gamma^5 d_b \, \bar{u}_a\gamma_\nu\gamma^5C\bar{d}^T_b\right)_r=&\left(Z_2^{-2} - \frac{4C_A^2-7}{32\pi^2\varepsilon C_A}g^2\right)u_a^T C\sigma^{\mu\nu}\gamma^5 d_b \, \bar{u}_a\gamma_\nu\gamma^5C\bar{d}^T_b\\
			&-\frac{3g^2}{32\pi^2\varepsilon} u_a^T C\sigma^{\mu\nu}\gamma^5 d_b \, \bar{u}_b\gamma_\nu\gamma^5C\bar{d}^T_a\\
			&-\frac{3iC_Ag^2}{32\pi^2\varepsilon}u_a^T C d_b \, \bar{u}_a\gamma^\mu C\bar{d}^T_b + \frac{3ig^2}{32\pi^2\varepsilon}u_a^T C d_b \, \bar{u}_b\gamma^\mu C\bar{d}^T_a 
		\end{split}
		\label{ren_tetra_2_a}
	\end{equation}
	\begin{equation}
		\begin{split}
			\left(u_a^T C\sigma^{\mu\nu}\gamma^5 d_b \,\bar{u}_b\gamma_\nu\gamma^5C\bar{d}^T_a\right)_r=&\left(Z_2^{-2} - \frac{4C_A^2-7}{32\pi^2\varepsilon C_A}g^2\right)u_a^T C\sigma^{\mu\nu}\gamma^5 d_b \,\bar{u}_b\gamma_\nu\gamma^5C\bar{d}^T_a\\
			&-\frac{3g^2}{32\pi^2\varepsilon} u_a^T C\sigma^{\mu\nu}\gamma^5 d_b \, \bar{u}_a\gamma_\nu\gamma^5C\bar{d}^T_b\\
			&+\frac{3iC_Ag^2}{32\pi^2\varepsilon}u_a^T C d_b \, \bar{u}_b\gamma^\mu C\bar{d}^T_a - \frac{3ig^2}{32\pi^2\varepsilon}u_a^T C d_b \, \bar{u}_a\gamma^\mu C\bar{d}^T_b \\
			&-\frac{1}{48\pi^2\varepsilon}\left(\bar{u}D^\alpha G_{\alpha\beta}\sigma^{\beta\mu}u - \bar{d}D^\alpha G_{\alpha\beta}\sigma^{\beta\mu}d\right)\\
			&-\frac{i}{16\pi^2\varepsilon}\left(\bar{u}D_\alpha G^{\alpha\mu}u - \bar{d}D_\alpha G^{\alpha\mu}d\right)
		\end{split}
		\label{ren_tetra_2_b}
	\end{equation}
	By replacing the $d$ quark with an $s$ quark, the renormalization of operator currents $\eta_6^\mu$ and $\eta_8^\mu$ can be obtained.

	\subsection{Correlation Functions}\label{apx1}
	\begin{equation}
		\begin{split}
			\Pi^V_{\eta_1}=&q^8\left[ \left(-\frac{79 g_s^2}{2654208 \pi ^8}-\frac{1}{18432 \pi ^6}\right) \log \left(-\frac{q^2}{\mu ^2}\right)-\frac{5 g_s^2 }{1327104 \pi ^8}\log ^2\left(-\frac{q^2}{\mu ^2}\right)\right] \\ 
			&+q^4\frac{ g_s^2  }{18432 \pi ^6}\log \left(-\frac{q^2}{\mu ^2}\right) 
			\langle  GG \rangle+-q^2\frac{ 1}{18 \pi ^2}\log \left(-\frac{q^2}{\mu ^2}\right) \left\langle  \bar{q}q \right\rangle ^2\\ 
			&+\frac{1}{12 \pi ^2}\log \left(-\frac{q^2}{\mu ^2}\right) \left\langle  \bar{q}q \right\rangle  \left\langle  \bar{q}Gq \right\rangle +\frac{1}{q^2}\left(\frac{5 g_s^2 }{864 \pi ^2}\langle  GG \rangle  \left\langle  \bar{q}q \right\rangle ^2-\frac{1}{48 \pi ^2}\left\langle  \bar{q}Gq \right\rangle ^2\right)\\ 
			&+\frac{1}{q^4}\left(-\frac{g_s^2}{576 \pi ^2} \langle  GG \rangle  \left\langle  \bar{q}q \right\rangle  \left\langle  \bar{q}Gq \right\rangle -\frac{32}{81} g_s^2 \left\langle  \bar{q}q \right\rangle ^4\right)
			\label{c_eta_1}
		\end{split}
		\notag
	\end{equation}

\begin{equation}
	\begin{split}
		\Pi^V_{\eta_2}=& q^8\left[\frac{65 g_s^2 }{1769472 \pi ^8}\log ^2\left(-\frac{q^2}{\mu ^2}\right)+\left(-\frac{297929 g_s^2}{557383680 \pi ^8}-\frac{1}{6144 \pi ^6}\right) \log \left(-\frac{q^2}{\mu ^2}\right)\right]\\ 
		&+-q^4 \frac{11 g_s }{18432 \pi ^6}\log \left(-\frac{q^2}{\mu ^2}\right) \langle  GG \rangle +-q^2 \frac{1}{6 \pi ^2}\log \left(-\frac{q^2}{\mu ^2}\right) \left\langle  \bar{q}q \right\rangle ^2\\ 
		&+\frac{1}{4 \pi ^2}\log \left(-\frac{q^2}{\mu ^2}\right) \left\langle  \bar{q}q \right\rangle  \left\langle  \bar{q}Gq \right\rangle +\frac{1}{q^2}\left(-\frac{5 g_s^2 }{864 \pi ^2}\langle  GG \rangle  \left\langle  \bar{q}q \right\rangle ^2-\frac{1}{16 \pi ^2}\left\langle  \bar{q}Gq \right\rangle ^2\right)\\ 
		&+\frac{1}{q^4}\left(\frac{g_s^2 }{576 \pi ^2}\langle  GG \rangle  \left\langle  \bar{q}q \right\rangle  \left\langle  \bar{q}Gq \right\rangle -\frac{32 g_s^2}{27} \left\langle  \bar{q}q \right\rangle ^4\right)
		\label{c_eta_2}
	\end{split}
\end{equation}

\begin{equation}
	\begin{split}
		\Pi^V_{\eta_3}=&q^8\left[\frac{5  g_s^2 }{884736 \pi ^8}\log ^2\left(-\frac{q^2}{\mu ^2}\right)+q^8 \left(-\frac{3503 g_s^2}{39813120 \pi ^8}-\frac{1}{36864 \pi ^6}\right) \log \left(-\frac{q^2}{\mu ^2}\right)\right]\\ 
		&+-q^4 \frac{g_s^2 }{18432 \pi ^6}\log \left(-\frac{q^2}{\mu ^2}\right) \langle  GG \rangle +-q^2 \frac{1}{36 \pi ^2} \log \left(-\frac{q^2}{\mu ^2}\right) \left\langle  \bar{q}q \right\rangle ^2\\ 
		&+\frac{1}{24 \pi ^2}\log \left(-\frac{q^2}{\mu ^2}\right) \left\langle  \bar{q}q \right\rangle  \left\langle  \bar{q}Gq \right\rangle +\frac{1}{q^2}\left(-\frac{5 g_s^2 }{864 \pi ^2}\langle  GG \rangle  \left\langle  \bar{q}q \right\rangle ^2-\frac{1}{96 \pi ^2}\left\langle  \bar{q}Gq \right\rangle ^2\right)\\ 
		&+\frac{1}{q^4}\left(\frac{g_s^2 }{576 \pi ^2}\langle  GG \rangle  \left\langle  \bar{q}q \right\rangle  \left\langle  \bar{q}Gq \right\rangle -\frac{16}{81} g_s^2 \left\langle  \bar{q}q \right\rangle ^4\right)
		\label{c_eta_3}
	\end{split}
\end{equation}

\begin{equation}
	\begin{split}
		\Pi^V_{\eta_4}=& q^8\left[\frac{5 g_s^2 }{1769472 \pi ^8}\log ^2\left(-\frac{q^2}{\mu ^2}\right)+ \left(-\frac{26021 g_s^2}{557383680 \pi ^8}-\frac{1}{12288 \pi ^6}\right) \log \left(-\frac{q^2}{\mu ^2}\right)\right]\\ 
		&+-q^4\frac{ g_s^2 }{18432 \pi ^6}\log \left(-\frac{q^2}{\mu ^2}\right) \langle  GG \rangle +-q^2\frac{1}{12 \pi ^2} \log \left(-\frac{q^2}{\mu ^2}\right) \left\langle  \bar{q}q \right\rangle ^2\\ 
		&+\frac{1}{8 \pi ^2}\log \left(-\frac{q^2}{\mu ^2}\right) \left\langle  \bar{q}q \right\rangle  \left\langle  \bar{q}Gq \right\rangle +\frac{1}{q^2}\left(\frac{5 g_s^2 }{864 \pi ^2}\langle  GG \rangle  \left\langle  \bar{q}q \right\rangle ^2-\frac{1}{32 \pi ^2}\left\langle  \bar{q}Gq \right\rangle ^2\right)\\ 
		&+\frac{1}{q^4}\left(-\frac{g_s^2 }{576 \pi ^2}\langle  GG \rangle  \left\langle  \bar{q}q \right\rangle  \left\langle  \bar{q}Gq \right\rangle-\frac{16}{27} g_s^2 \left\langle  \bar{q}q \right\rangle ^4\right)
		\label{c_eta_4}
	\end{split}
\end{equation}

\begin{equation}
	\begin{split}
		\Pi^V_{\eta_5}=&q^8 \left[\left(-\frac{2243 g_s^2}{79626240 \pi ^8}-\frac{1}{18432 \pi ^6}\right) \log \left(-\frac{q^2}{\mu ^2}\right)-\frac{7 g_s^2 }{1769472 \pi ^8}\log ^2\left(-\frac{q^2}{\mu ^2}\right)\right]\\
		&+q^6\frac{17  m_s^2 }{7680 \pi ^6}\log \left(-\frac{q^2}{\mu ^2}\right)+q^4 \log \left(-\frac{q^2}{\mu ^2}\right) \left(-\frac{m_s \left\langle  \bar{s}s\text{}\right\rangle }{48 \pi ^4}+\frac{m_s \left\langle  \bar{q}q \right\rangle }{96 \pi ^4}+\frac{g_s^2 \langle  GG \rangle }{18432 \pi ^6}\right)\\ 
		&+q^2 \log \left(-\frac{q^2}{\mu ^2}\right) \left(\frac{m_s \left\langle  \bar{s}\text{G}s\text{}\right\rangle }{96 \pi ^4}-\frac{\left\langle  \bar{q}q \right\rangle  \left\langle  \bar{s}s\text{}\right\rangle }{9 \pi ^2}+\frac{\left\langle  \bar{s}s\text{}\right\rangle ^2}{36 \pi ^2}-\frac{m_s \left\langle  \bar{q}Gq \right\rangle }{48 \pi ^4}+\frac{\left\langle  \bar{q}q \right\rangle ^2}{36 \pi ^2}-\frac{g_s^2 m_s^2 \langle  GG \rangle }{4608 \pi ^6}\right)\\ 
		&+\log \left(-\frac{q^2}{\mu ^2}\right) \left(\frac{\left\langle  \bar{q}q \right\rangle  \left\langle  \bar{s}\text{G}s\text{}\right\rangle }{12 \pi ^2}-\frac{\left\langle  \bar{s}s\text{}\right\rangle  \left\langle  \bar{s}\text{G}s\text{}\right\rangle }{24 \pi ^2}+\frac{\left\langle  \bar{s}s\text{}\right\rangle  \left\langle  \bar{q}Gq \right\rangle }{12 \pi ^2}+\frac{m_s^2 \left\langle  \bar{q}q \right\rangle  \left\langle  \bar{s}s\text{}\right\rangle }{2 \pi ^2}-\frac{m_s^2 \left\langle  \bar{s}s\text{}\right\rangle ^2}{24 \pi ^2}\right.\\
		&\left.-\frac{g_s^2 m_s \langle  GG \rangle  \left\langle  \bar{q}q \right\rangle }{256 \pi ^4}-\frac{\left\langle  \bar{q}q \right\rangle  \left\langle  \bar{q}Gq \right\rangle }{24 \pi ^2}+\frac{m_s^2 \left\langle  \bar{q}q \right\rangle ^2}{6 \pi ^2}\right)+\frac{1}{q^2}\left(-\frac{\left\langle  \bar{q}Gq \right\rangle  \left\langle  \bar{s}\text{G}s\text{}\right\rangle }{24 \pi ^2}-\frac{m_s^2 \left\langle  \bar{q}q \right\rangle  \left\langle  \bar{s}\text{G}s\text{}\right\rangle }{6 \pi ^2}\right.\\
		&\left.+\frac{\left\langle  \bar{s}\text{G}s\text{}\right\rangle ^2}{96 \pi ^2}\right.\left.+\frac{5 g_s^2 \langle  GG \rangle  \left\langle  \bar{q}q \right\rangle  \left\langle  \bar{s}s\text{}\right\rangle }{864 \pi ^2}-\frac{m_s^2 \left\langle  \bar{s}s\text{}\right\rangle  \left\langle  \bar{q}Gq \right\rangle }{4 \pi ^2}-\frac{4}{9} m_s \left\langle  \bar{q}q \right\rangle  \left\langle  \bar{s}s\text{}\right\rangle ^2-\frac{2}{3} m_s \left\langle  \bar{q}q \right\rangle ^2 \left\langle  \bar{s}s\text{}\right\rangle\right.\\
		&\left. +\frac{5 g_s^2 m_s \langle  GG \rangle  \left\langle  \bar{q}Gq \right\rangle }{4608 \pi ^4}+\frac{\left\langle  \bar{q}Gq \right\rangle ^2}{96 \pi ^2}\right)+\frac{1}{q^4}\left(-\frac{g_s^2 \langle  GG \rangle  \left\langle  \bar{q}q \right\rangle  \left\langle  \bar{s}\text{G}s\text{}\right\rangle }{1152 \pi ^2}-\frac{m_s^2 \left\langle  \bar{q}Gq \right\rangle  \left\langle  \bar{s}\text{G}s\text{}\right\rangle }{24 \pi ^2}\right.\\
		&\left.+\frac{2}{9} m_s \left\langle  \bar{q}q \right\rangle ^2 \left\langle  \bar{s}\text{G}s\text{}\right\rangle -\frac{1}{9} m_s \left\langle  \bar{q}q \right\rangle  \left\langle  \bar{s}s\text{}\right\rangle  \left\langle  \bar{s}\text{G}s\text{}\right\rangle -\frac{g_s^2 \langle  GG \rangle  \left\langle  \bar{s}s\text{}\right\rangle  \left\langle  \bar{q}Gq \right\rangle }{1152 \pi ^2}-\frac{32}{81} g_s^2 \left\langle  \bar{q}q \right\rangle ^2 \left\langle  \bar{s}s\text{}\right\rangle ^2\right.\\
		&\left.-\frac{1}{9} m_s \left\langle  \bar{s}s\text{}\right\rangle ^2 \left\langle  \bar{q}Gq \right\rangle +\frac{5}{9} m_s \left\langle  \bar{q}q \right\rangle  \left\langle  \bar{s}s\text{}\right\rangle  \left\langle  \bar{q}Gq \right\rangle +\frac{m_s^2 \left\langle  \bar{q}Gq \right\rangle ^2}{24 \pi ^2}\right)
		\label{c_eta_5}
	\end{split}
\end{equation}

\begin{equation}
	\begin{split}
		\Pi^V_{\eta_6}=&q^8\left[ \frac{65 g_s^2 }{1769472 \pi ^8}\log ^2\left(-\frac{q^2}{\mu ^2}\right)+ \left(-\frac{297929 g_s^2}{557383680 \pi ^8}-\frac{1}{6144 \pi ^6}\right) \log \left(-\frac{q^2}{\mu ^2}\right)\right]\\ 
		&+ q^6\frac{17 m_s^2 }{2560 \pi ^6}\log \left(-\frac{q^2}{\mu ^2}\right)+q^4 \log \left(-\frac{q^2}{\mu ^2}\right) \left(-\frac{m_s \left\langle  \bar{s}s\text{}\right\rangle }{16 \pi ^4}+\frac{m_s \left\langle  \bar{q}q \right\rangle }{32 \pi ^4}-\frac{11 g_s^2 \langle  GG \rangle }{18432 \pi ^6}\right)\\
		&+q^2 \log \left(-\frac{q^2}{\mu ^2}\right) \left(\frac{m_s \left\langle  \bar{s}\text{G}s\text{}\right\rangle }{32 \pi ^4}-\frac{\left\langle  \bar{q}q \right\rangle  \left\langle  \bar{s}s\text{}\right\rangle }{3 \pi ^2}+\frac{\left\langle  \bar{s}s\text{}\right\rangle ^2}{12 \pi ^2}-\frac{m_s \left\langle  \bar{q}Gq \right\rangle }{16 \pi ^4}+\frac{\left\langle  \bar{q}q \right\rangle ^2}{12 \pi ^2}+\frac{109 g_s^2 m_s^2 \langle  GG \rangle }{18432 \pi ^6}\right)\\		
		&+\log \left(-\frac{q^2}{\mu ^2}\right) \left(\frac{\left\langle  \bar{q}q \right\rangle  \left\langle  \bar{s}\text{G}s\text{}\right\rangle }{4 \pi ^2}-\frac{\left\langle  \bar{s}s\text{}\right\rangle  \left\langle  \bar{s}\text{G}s\text{}\right\rangle }{8 \pi ^2}-\frac{5 g_s^2 m_s \langle  GG \rangle  \left\langle  \bar{s}s\text{}\right\rangle }{256 \pi ^4}+\frac{\left\langle  \bar{s}s\text{}\right\rangle  \left\langle  \bar{q}Gq \right\rangle }{4 \pi ^2}+\frac{3 m_s^2 \left\langle  \bar{q}q \right\rangle  \left\langle  \bar{s}s\text{}\right\rangle }{2 \pi ^2}\right.\\
		&\left.-\frac{m_s^2 \left\langle  \bar{s}s\text{}\right\rangle ^2}{8 \pi ^2}+\frac{3 g_s^2 m_s \langle  GG \rangle  \left\langle  \bar{q}q \right\rangle }{128 \pi ^4}-\frac{\left\langle  \bar{q}q \right\rangle  \left\langle  \bar{q}Gq \right\rangle }{8 \pi ^2}+\frac{m_s^2 \left\langle  \bar{q}q \right\rangle ^2}{2 \pi ^2}\right)+\frac{1}{q^2}\left(\frac{25 g_s^2 m_s \langle  GG \rangle  \left\langle  \bar{s}\text{G}s\text{}\right\rangle }{4608 \pi ^4}\right.\\
		&\left.-\frac{\left\langle  \bar{q}Gq \right\rangle  \left\langle  \bar{s}\text{G}s\text{}\right\rangle }{8 \pi ^2}-\frac{m_s^2 \left\langle  \bar{q}q \right\rangle  \left\langle  \bar{s}\text{G}s\text{}\right\rangle }{2 \pi ^2}+\frac{\left\langle  \bar{s}\text{G}s\text{}\right\rangle ^2}{32 \pi ^2}-\frac{5 g_s^2 \langle  GG \rangle  \left\langle  \bar{q}q \right\rangle  \left\langle  \bar{s}s\text{}\right\rangle }{144 \pi ^2}+\frac{25 g_s^2 \langle  GG \rangle  \left\langle  \bar{s}s\text{}\right\rangle ^2}{1728 \pi ^2}\right.\\
		&\left.-\frac{3 m_s^2 \left\langle  \bar{s}s\text{}\right\rangle  \left\langle  \bar{q}Gq \right\rangle }{4 \pi ^2}-\frac{4}{3} m_s \left\langle  \bar{q}q \right\rangle  \left\langle  \bar{s}s\text{}\right\rangle ^2-2 m_s \left\langle  \bar{q}q \right\rangle ^2 \left\langle  \bar{s}s\text{}\right\rangle -\frac{5 g_s^2 m_s \langle  GG \rangle  \left\langle  \bar{q}Gq \right\rangle }{768 \pi ^4}+\frac{25 g_s^2 \langle  GG \rangle  \left\langle  \bar{q}q \right\rangle ^2}{1728 \pi ^2}\right.
	\end{split}
\end{equation}


\begin{equation}
	\begin{split}
		&\left.+\frac{\left\langle  \bar{q}Gq \right\rangle ^2}{32 \pi ^2}\right)+\frac{1}{q^4}\left(\frac{g_s^2 \langle  GG \rangle  \left\langle  \bar{q}q \right\rangle  \left\langle  \bar{s}\text{G}s\text{}\right\rangle }{192 \pi ^2}-\frac{5 g_s^2 \langle  GG \rangle  \left\langle  \bar{s}s\text{}\right\rangle  \left\langle  \bar{s}\text{G}s\text{}\right\rangle }{1152 \pi ^2}-\frac{m_s^2 \left\langle  \bar{q}Gq \right\rangle  \left\langle  \bar{s}\text{G}s\text{}\right\rangle }{8 \pi ^2}\right.\\
		&\left.+\frac{2}{3} m_s \left\langle  \bar{q}q \right\rangle ^2 \left\langle  \bar{s}\text{G}s\text{}\right\rangle -\frac{1}{3} m_s \left\langle  \bar{q}q \right\rangle  \left\langle  \bar{s}s\text{}\right\rangle  \left\langle  \bar{s}\text{G}s\text{}\right\rangle -\frac{5 g_s^2 m_s^2 \langle  GG \rangle  \left\langle  \bar{s}s\text{}\right\rangle ^2}{1152 \pi ^2}+\frac{g_s^2 \langle  GG \rangle  \left\langle  \bar{s}s\text{}\right\rangle  \left\langle  \bar{q}Gq \right\rangle }{192 \pi ^2}\right.\\
		&\left.-\frac{32}{27} g_s^2 \left\langle  \bar{q}q \right\rangle ^2 \left\langle  \bar{s}s\text{}\right\rangle ^2-\frac{1}{3} m_s \left\langle  \bar{s}s\text{}\right\rangle ^2 \left\langle  \bar{q}Gq \right\rangle +\frac{5}{3} m_s \left\langle  \bar{q}q \right\rangle  \left\langle  \bar{s}s\text{}\right\rangle  \left\langle  \bar{q}Gq \right\rangle -\frac{5 g_s^2 \langle  GG \rangle  \left\langle  \bar{q}q \right\rangle  \left\langle  \bar{q}Gq \right\rangle }{1152 \pi ^2}\right.\\
		&\left.+\frac{m_s^2 \left\langle  \bar{q}Gq \right\rangle ^2}{8 \pi ^2}\right)
		\label{c_eta_6}
	\end{split}
\end{equation}

\begin{equation}
	\begin{split}
		\Pi^V_{\eta_7}=& q^8\left[\frac{5 g_s^2 }{1769472 \pi ^8}\log ^2\left(-\frac{q^2}{\mu ^2}\right)+ \left(-\frac{3503 g_s^2}{79626240 \pi ^8}-\frac{1}{36864 \pi ^6}\right) \log \left(-\frac{q^2}{\mu ^2}\right)\right]\\ 
		&+q^6 \frac{17 m_s^2 }{15360 \pi ^6}\log \left(-\frac{q^2}{\mu ^2}\right)+q^4 \log \left(-\frac{q^2}{\mu ^2}\right) \left(-\frac{m_s \left\langle  \bar{s}s\text{}\right\rangle }{96 \pi ^4}+\frac{m_s \left\langle  \bar{q}q \right\rangle }{192 \pi ^4}-\frac{g_s^2 \langle  GG \rangle }{18432 \pi ^6}\right)\\
		&+q^2 \log \left(-\frac{q^2}{\mu ^2}\right) \left(\frac{m_s \left\langle  \bar{s}\text{G}s\text{}\right\rangle }{192 \pi ^4}-\frac{\left\langle  \bar{q}q \right\rangle  \left\langle  \bar{s}s\text{}\right\rangle }{18 \pi ^2}+\frac{\left\langle  \bar{s}s\text{}\right\rangle ^2}{72 \pi ^2}-\frac{m_s \left\langle  \bar{q}Gq \right\rangle }{96 \pi ^4}+\frac{\left\langle  \bar{q}q \right\rangle ^2}{72 \pi ^2}+\frac{g_s^2 m_s^2 \langle  GG \rangle }{4608 \pi ^6}\right)\\ 
		&+\log \left(-\frac{q^2}{\mu ^2}\right) \left(\frac{\left\langle  \bar{q}q \right\rangle  \left\langle  \bar{s}\text{G}s\text{}\right\rangle }{24 \pi ^2}-\frac{\left\langle  \bar{s}s\text{}\right\rangle  \left\langle  \bar{s}\text{G}s\text{}\right\rangle }{48 \pi ^2}+\frac{\left\langle  \bar{s}s\text{}\right\rangle  \left\langle  \bar{q}Gq \right\rangle }{24 \pi ^2}+\frac{m_s^2 \left\langle  \bar{q}q \right\rangle  \left\langle  \bar{s}s\text{}\right\rangle }{4 \pi ^2}-\frac{m_s^2 \left\langle  \bar{s}s\text{}\right\rangle ^2}{48 \pi ^2}\right.\\
		&\left.+\frac{g_s^2 m_s \langle  GG \rangle  \left\langle  \bar{q}q \right\rangle }{256 \pi ^4}-\frac{\left\langle  \bar{q}q \right\rangle  \left\langle  \bar{q}Gq \right\rangle }{48 \pi ^2}+\frac{m_s^2 \left\langle  \bar{q}q \right\rangle ^2}{12 \pi ^2}\right)+\frac{1}{q^2}\left(-\frac{\left\langle  \bar{q}Gq \right\rangle  \left\langle  \bar{s}\text{G}s\text{}\right\rangle }{48 \pi ^2}-\frac{m_s^2 \left\langle  \bar{q}q \right\rangle  \left\langle  \bar{s}\text{G}s\text{}\right\rangle }{12 \pi ^2}\right.\\
		&\left.+\frac{\left\langle  \bar{s}\text{G}s\text{}\right\rangle ^2}{192 \pi ^2}+\frac{5 g_s^2 \langle  GG \rangle  \left\langle  \bar{q}q \right\rangle  \left\langle  \bar{s}s\text{}\right\rangle }{864 \pi ^2}-\frac{m_s^2 \left\langle  \bar{s}s\text{}\right\rangle  \left\langle  \bar{q}Gq \right\rangle }{8 \pi ^2}-\frac{2}{9} m_s \left\langle  \bar{q}q \right\rangle  \left\langle  \bar{s}s\text{}\right\rangle ^2-\frac{1}{3} m_s \left\langle  \bar{q}q \right\rangle ^2 \left\langle  \bar{s}s\text{}\right\rangle\right.\\
		&\left. -\frac{5 g_s^2 m_s \langle  GG \rangle  \left\langle  \bar{q}Gq \right\rangle }{4608 \pi ^4}+\frac{\left\langle  \bar{q}Gq \right\rangle ^2}{192 \pi ^2}\right)+\frac{1}{q^4}\left(-\frac{m_s^2 \left\langle  \bar{q}Gq \right\rangle  \left\langle  \bar{s}\text{G}s\text{}\right\rangle }{48 \pi ^2}+\frac{1}{9} m_s \left\langle  \bar{q}q \right\rangle ^2 \left\langle  \bar{s}\text{G}s\text{}\right\rangle \right.\\
		&\left.-\frac{1}{18} m_s \left\langle  \bar{q}q \right\rangle  \left\langle  \bar{s}s\text{}\right\rangle  \left\langle  \bar{s}\text{G}s\text{}\right\rangle +\frac{g_s^2 \langle  GG \rangle  \left\langle  \bar{s}s\text{}\right\rangle  \left\langle  \bar{q}Gq \right\rangle }{1152 \pi ^2}-\frac{16}{81} g_s^2 \left\langle  \bar{q}q \right\rangle ^2 \left\langle  \bar{s}s\text{}\right\rangle ^2-\frac{1}{18} m_s \left\langle  \bar{s}s\text{}\right\rangle ^2 \left\langle  \bar{q}Gq \right\rangle \right.\\
		&\left.+\frac{5}{18} m_s \left\langle  \bar{q}q \right\rangle  \left\langle  \bar{s}s\text{}\right\rangle  \left\langle  \bar{q}Gq \right\rangle +\frac{g_s^2 \langle  GG \rangle  \left\langle  \bar{q}q \right\rangle  \left\langle  \bar{q}Gq \right\rangle }{1152 \pi ^2}+\frac{m_s^2 \left\langle  \bar{q}Gq \right\rangle ^2}{48 \pi ^2}\right)
		\label{c_eta_7}
	\end{split}
\end{equation}

\begin{equation}
	\begin{split}
		\Pi^V_{\eta_8}=&q^8 \left[\frac{5 g_s^2 }{1769472 \pi ^8}\log ^2\left(-\frac{q^2}{\mu ^2}\right)+ \left(-\frac{26021 g_s^2}{557383680 \pi ^8}-\frac{1}{12288 \pi ^6}\right) \log \left(-\frac{q^2}{\mu ^2}\right)\right]\\ 
		&+ q^6\frac{17 m_s^2 }{5120 \pi ^6}\log \left(-\frac{q^2}{\mu ^2}\right)+q^4 \log \left(-\frac{q^2}{\mu ^2}\right) \left(-\frac{m_s \left\langle  \bar{s}s\text{}\right\rangle }{32 \pi ^4}+\frac{m_s \left\langle  \bar{q}q \right\rangle }{64 \pi ^4}-\frac{g_s^2 \langle  GG \rangle }{18432 \pi ^6}\right)\\
		&+q^2 \log \left(-\frac{q^2}{\mu ^2}\right) \left(\frac{m_s \left\langle  \bar{s}\text{G}s\text{}\right\rangle }{64 \pi ^4}-\frac{\left\langle  \bar{q}q \right\rangle  \left\langle  \bar{s}s\text{}\right\rangle }{6 \pi ^2}+\frac{\left\langle  \bar{s}s\text{}\right\rangle ^2}{24 \pi ^2}-\frac{m_s \left\langle  \bar{q}Gq \right\rangle }{32 \pi ^4}+\frac{\left\langle  \bar{q}q \right\rangle ^2}{24 \pi ^2}+\frac{17 g_s^2 m_s^2 \langle  GG \rangle }{18432 \pi ^6}\right)\\ 
		&+\log \left(-\frac{q^2}{\mu ^2}\right) \left(\frac{\left\langle  \bar{q}q \right\rangle  \left\langle  \bar{s}\text{G}s\text{}\right\rangle }{8 \pi ^2}-\frac{\left\langle  \bar{s}s\text{}\right\rangle  \left\langle  \bar{s}\text{G}s\text{}\right\rangle }{16 \pi ^2}-\frac{g_s^2 m_s \langle  GG \rangle  \left\langle  \bar{s}s\text{}\right\rangle }{256 \pi ^4}+\frac{\left\langle  \bar{s}s\text{}\right\rangle  \left\langle  \bar{q}Gq \right\rangle }{8 \pi ^2}+\frac{3 m_s^2 \left\langle  \bar{q}q \right\rangle  \left\langle  \bar{s}s\text{}\right\rangle }{4 \pi ^2}\right.\\
		&\left.-\frac{m_s^2 \left\langle  \bar{s}s\text{}\right\rangle ^2}{16 \pi ^2}-\frac{\left\langle  \bar{q}q \right\rangle  \left\langle  \bar{q}Gq \right\rangle }{16 \pi ^2}+\frac{m_s^2 \left\langle  \bar{q}q \right\rangle ^2}{4 \pi ^2}\right)+\frac{1}{q^2}\left(\frac{5 g_s^2 m_s \langle  GG \rangle  \left\langle  \bar{s}\text{G}s\text{}\right\rangle }{4608 \pi ^4}-\frac{\left\langle  \bar{q}Gq \right\rangle  \left\langle  \bar{s}\text{G}s\text{}\right\rangle }{16 \pi ^2}\right.\\
		&\left.-\frac{m_s^2 \left\langle  \bar{q}q \right\rangle  \left\langle  \bar{s}\text{G}s\text{}\right\rangle }{4 \pi ^2}+\frac{\left\langle  \bar{s}\text{G}s\text{}\right\rangle ^2}{64 \pi ^2}+\frac{5 g_s^2 \langle  GG \rangle  \left\langle  \bar{s}s\text{}\right\rangle ^2}{1728 \pi ^2}-\frac{3 m_s^2 \left\langle  \bar{s}s\text{}\right\rangle  \left\langle  \bar{q}Gq \right\rangle }{8 \pi ^2}-\frac{2}{3} m_s \left\langle  \bar{q}q \right\rangle  \left\langle  \bar{s}s\text{}\right\rangle ^2\right.\\
		&\left.-m_s \left\langle  \bar{q}q \right\rangle ^2 \left\langle  \bar{s}s\text{}\right\rangle +\frac{5 g_s^2 \langle  GG \rangle  \left\langle  \bar{q}q \right\rangle ^2}{1728 \pi ^2}+\frac{\left\langle  \bar{q}Gq \right\rangle ^2}{64 \pi ^2}\right)+\frac{1}{q^4}\left(-\frac{g_s^2 \langle  GG \rangle  \left\langle  \bar{s}s\text{}\right\rangle  \left\langle  \bar{s}\text{G}s\text{}\right\rangle }{1152 \pi ^2}\right.\\
		&\left.-\frac{m_s^2 \left\langle  \bar{q}Gq \right\rangle  \left\langle  \bar{s}\text{G}s\text{}\right\rangle }{16 \pi ^2}+\frac{1}{3} m_s \left\langle  \bar{q}q \right\rangle ^2 \left\langle  \bar{s}\text{G}s\text{}\right\rangle -\frac{1}{6} m_s \left\langle  \bar{q}q \right\rangle  \left\langle  \bar{s}s\text{}\right\rangle  \left\langle  \bar{s}\text{G}s\text{}\right\rangle -\frac{g_s^2 m_s^2 \langle  GG \rangle  \left\langle  \bar{s}s\text{}\right\rangle ^2}{1152 \pi ^2}\right.\\
		&\left.-\frac{16}{27} g_s^2 \left\langle  \bar{q}q \right\rangle ^2 \left\langle  \bar{s}s\text{}\right\rangle ^2-\frac{1}{6} m_s \left\langle  \bar{s}s\text{}\right\rangle ^2 \left\langle  \bar{q}Gq \right\rangle +\frac{5}{6} m_s \left\langle  \bar{q}q \right\rangle  \left\langle  \bar{s}s\text{}\right\rangle  \left\langle  \bar{q}Gq \right\rangle -\frac{g_s^2 \langle  GG \rangle  \left\langle  \bar{q}q \right\rangle  \left\langle  \bar{q}Gq \right\rangle }{1152 \pi ^2}\right.\\
		&\left.+\frac{m_s^2 \left\langle  \bar{q}Gq \right\rangle ^2}{16 \pi ^2}\right)
		\label{c_eta_8}
	\end{split}
\end{equation}

\begin{equation}
	\begin{split}
		\Pi^V_{J_1}=& q^8 \left[\frac{25 g_s^2 }{14155776 \pi ^8}\log ^2\left(-\frac{q^2}{\mu ^2}\right)+ \left(-\frac{41221 g_s^2}{1114767360 \pi ^8}-\frac{11}{1179648 \pi ^6}\right) \log \left(-\frac{q^2}{\mu ^2}\right)\right]\\ 
		&+-q^4\frac{ g_s^2 }{32768 \pi ^6}\log \left(-\frac{q^2}{\mu ^2}\right) \langle  GG \rangle +q^2 \frac{1}{128 \pi ^2}\log \left(-\frac{q^2}{\mu ^2}\right) \left\langle  \bar{q}q\text{}\right\rangle ^2\\ 
		&+-\frac{3 }{256 \pi ^2}\log \left(-\frac{q^2}{\mu ^2}\right) \left\langle  \bar{q}q\text{}\right\rangle  \left\langle  \bar{q}\text{G}q\text{}\right\rangle +\frac{1}{q^2}\left(\frac{3 }{1024 \pi ^2}\left\langle  \bar{q}\text{G}q\text{}\right\rangle ^2+\frac{5 g_s^2}{1536 \pi ^2} \langle  GG \rangle  \left\langle  \bar{q}q\text{}\right\rangle ^2\right)\\ 
		&+\frac{1}{q^4}\frac{g_s^2 }{3072 \pi ^2 }\langle  GG \rangle  \left\langle  \bar{q}q\text{}\right\rangle  \left\langle  \bar{q}\text{G}q\text{}\right\rangle 
		\label{c_mol_1}
	\end{split}
\end{equation}

\begin{equation}
	\begin{split}
		\Pi^V_{J_2}=&q^8 \left[\left(-\frac{77 g_s^2}{2764800 \pi ^8}-\frac{1}{30720 \pi ^6}\right) \log \left(-\frac{q^2}{\mu ^2}\right)-\frac{g_s^2 }{829440 \pi ^8}\log ^2\left(-\frac{q^2}{\mu ^2}\right)\right]\\ 
		&+q^4\frac{ g_s^2 }{3072 \pi ^6}\log \left(-\frac{q^2}{\mu ^2}\right) \langle  GG \rangle +q^2 \left[ \left(-\frac{5 \gamma  g_s^2 }{216 \pi ^4}-\frac{5 g_s^2 }{1296 \pi ^4}\right)\log \left(-\frac{q^2}{\mu ^2}\right)\right.\\
		&\left.-\frac{5 g_s^2 }{216 \pi ^4}\log ^2\left(-\frac{q^2}{\mu ^2}\right) \right]\left\langle  \bar{q}q \right\rangle ^2+\frac{1}{48 \pi ^2}\log \left(-\frac{q^2}{\mu ^2}\right) \left\langle  \bar{q}q \right\rangle  \left\langle  \bar{q}Gq \right\rangle \\
		&+\frac{1}{q^2}\left(-\frac{g_s^2 }{864 \pi ^2}\langle  GG \rangle  \left\langle  \bar{q}q \right\rangle ^2-\frac{1}{192 \pi ^2}\left\langle  \bar{q}Gq \right\rangle ^2\right)+\frac{1}{q^4}\frac{g_s^2 }{1728 \pi ^2 }\langle  GG \rangle  \left\langle  \bar{q}q \right\rangle  \left\langle  \bar{q}Gq \right\rangle 
		\label{c_mol_2}
	\end{split}
\end{equation}

	\bibliographystyle{JHEP}
	\bibliography{refs.bib}
	
\end{document}